\documentstyle[aps,epsf,psfig,prl,twocolumn,amsfonts,floats]{revtex}

\newcommand{\nm}{nm}   
\newcommand{\um}{$\mu$m}
\newcommand{\ps}{$\mbox{s}^{-1}$}

\newcommand{\umumps}{$\mbox{\um}^2\mbox{s}^{-1}$}

\newcommand{\pon} {{p^{\mbox{\scriptsize on}} }}
\newcommand{\poff}{{p^{\mbox{\scriptsize off}} }}
\newcommand{\mts} {microtubules}
\newcommand{\mt} {microtubule}

\begin{document}
\title{Dynamic concentration of motors in microtubule arrays} \address
{Fran{\c c}ois N{\'e}d{\'e}lec$^\star$,
  Thomas Surrey$^\star$ and Anthony Maggs$^\dagger$\\
  $^\star$EMBL, Cell Biology, Meyerhofstrasse 1,
  69115 Heidelberg, Germany\\
  $^\dagger$ESPCI, PCT, 10 Rue Vauquelin, 75005 Paris, France}
\address{\begin{minipage}{5.5in}
\begin{abstract}
  We present experimental and theoretical studies of the dynamics of
  molecular motors in microtubule arrays and asters. By solving a
  convection-diffusion equation we find that the density profile of
  motors in a two-dimensional aster is characterized by continuously
  varying exponents. Simulations are used to verify the assumptions of
  the continuum model.  We observe the concentration profiles of
  kinesin moving in quasi two-dimensional artificial asters by
  fluorescent microscopy and compare with our theoretical results.
\end{abstract}
\end{minipage}
\date{\today} }\maketitle

The cytoskeleton is a network of polymers essential for the dynamic
organization of many eukaryotic cells. Its function depends not only
on protein fibers, but also on many accessory components~\cite{CMP}.
Among these are motor proteins that reversibly bind to and walk along
the surface of cytoskeletal polymers, consuming ATP as a source of
energy.  A natural consequence of this directed movement in organized
fiber arrays is an non-uniform spatial distribution of the motors.  In
vivo \mts\ are often observed in radial arrays, or asters, where all \mt
``minus'' ends are at the center, and the ``plus'' ends are radiating
outward. Aster of opposite polarity, in which kinesin moves inward can
also be formed in-vitro~\cite{NedelecSurrey00}.

\begin{figure}[htb]  
  \psfig{figure=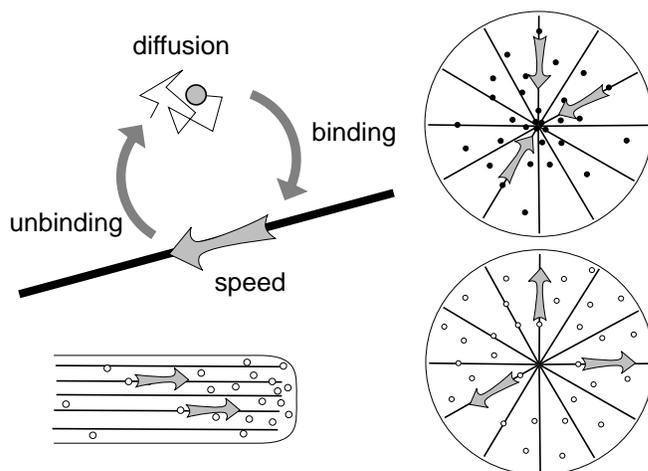,width=8.6cm}
\caption{ 
  In the presence of a \mt\ array motors can move by free diffusion in
  solution or by directed motion on \mts.  Movement of motors in an
  aster can lead to accumulation, if the motor moves inward (top
  right), or depletion, if the motor moves outward (bottom right).
  Accumulation also occurs in oriented parallel \mt\ arrays (bottom
  left). }
\label{FIGIntro}
\end{figure}

In this letter we present quantitative experiments on a quasi two
dimensional aster, fig.~(\ref{FIGIntro}), formed from long \mts\
polymerized between closely separated cover slips.  We analyze the
results of these experiments using a 2D transport diffusion equation.
The theory contains a number of approximations, due to angular
averaging and projection from 3 to 2 dimensions. We verify that no
major quantitative error is introduced by performing simulations in
the full confined geometry.  We also consider, theoretically, the case
of one and three dimensions: The aster in one dimension corresponds to
a tube in which \mts\ are all oriented in the same direction.  This is
the case for example in the axons of nerve cells.  As distinct motors
move toward the ``plus'' or the ``minus'' end, we consider both cases
of inward/outward directed motion.

Consider $N$ immobile straight \mts\ radially arranged in the available
volume. Molecular motors are present which can exist in two different
states, either attached to a filament, or detached.  Unattached motors
diffuse freely, with a diffusion constant $D$.  Attached motors move
on their filament (radially in the aster geometry) at a velocity $v$.
Positive values of $v$ corresponds to outward movement.  Transitions
between the two states are stochastic:  The motor spontaneously
detaches from the \mt\ at an unbinding rate $\poff$ ($s^{-1}$).  Far from
saturation, the number of binding events per second is proportional to
the local concentration of free motors, and to the number of available
binding sites on the \mts.  If the concentration of free motors is
expressed in molecules per cubic micrometers, and the available
``quantity'' of \mts\ in micrometers, the constant of proportionality
$\pon$ has the dimension of a diffusion
constant\cite{FosterGilbert98,HackneyNature95}.

Let $b(r)$ and $f(r)$ be the concentrations of bound and free motors,
respectively, at distance $r$ from the center, averaged over all
angles.  Bound motors move radially at speed $v$, and create a
convective flux $J_b=v b$.  Unbound motors diffuse freely, creating a
radial flux $J_f=-Ddf/dr$.  If $S$ is the surface area at distance
$r$, there are $\poff Sb\, dr$ release events in the volume between
the radii $[r, r+dr]$, and $\pon Nf\, dr$ attachments per second.  We
therefore obtain the coupled kinetic equations:
\begin{eqnarray}
\label{EQNdynamic}
{{\partial b}\over {\partial t}} &=& -\poff b + 
\frac{\pon N}{S}f -{{1}\over {S}}\frac{\partial}{\partial r}(J_b S) \nonumber \\ 
\frac{\partial f}{\partial t} &=& +\poff b -
\frac{\pon N}{S}f - \frac{1}{S}\frac{\partial }{\partial r}(J_f S)
\end{eqnarray}
To find the steady state, we use the fact that the net flux $J_b +
J_f$ is zero, implying $b=Dv^{-1}(df/dr)$.  Substituting this result
into equation (\ref{EQNdynamic}), and denoting $f'=df/dr$ and
$f''=d^2f/{dr}^2$, we find in $d$ dimensions
\begin{eqnarray}
0  &=&  f'' +  \left( \frac{1}{\alpha}+\frac{d-1}{r}\right) f'  
-\frac{1}{r^{d-1} \beta_d}f \nonumber \\
b &=& \gamma \,f'
\end{eqnarray}
The physical parameters in the problem have been reduced to
\begin{eqnarray}
\alpha &=& v/\poff \nonumber \\ 
\beta_1= S_1 D/p^{on}N ,
\quad   \beta_2 &=& {2\pi h D}/{N\pon},
\quad 
\beta_3= 4 \pi D/p^{on} N \nonumber
\\
\gamma &=& D/v.
\end{eqnarray}
where $S_1$ is the area of the tube in one dimension and $h$ is the
sample thickness in a quasi two-dimensional geometry.  $\alpha$ is the
average distance that a motor moves on a \mt, before detaching.
$\beta_d$ characterizes the geometry of the aster, and $\gamma$
determines the relative concentration of bound to free motors.  Note
that for inward movement, $\alpha$ and $\gamma$ have negative values,
while for outward movement all parameters are positive.

For the motor protein kinesin, experimental data provide values for
the parameters in the model.  The walking speed of kinesin without
load is $v=0.8$~\um\ps\ \cite{HowardVale89,SvobodaBlock94}.  The
unbinding rate $\poff$ is obtained from the average distance that
kinesin moves before detaching.  Measured average run length, $\alpha
\, = \, v/\poff$, are for kinesin in the range $0.4-1.5$~\um\ 
\cite{BlockSchnapp90,ValeYanagida96,CoyHoward99} ; we use $\poff =
1$~\ps.  Direct chemical measurements of $\poff$
\cite{HackneyNature95} agree with this value.  The binding rate $\pon$
of the kinesin construct used in our experiment has not been directly
measured.  To estimate it, we assume that interaction between \mts\ and
motors is diffusion limited \cite{Hackney95}: Measurements with
kinesin's soluble dimeric motor domain \cite{HackneyNature95} provide
a value of $\pon_{kin} = 7.3$~\umumps, and its diffusion constant is
$50$~\umumps\ \cite{HackneySuhan92,HuangHackney94}.  For single
kinesin adsorbed on beads \cite{CoyHoward99}, the equilibrium constant
for the binding convection (equal to $\pon/\poff$), provides
$\pon_{bead} = 0.25$~\umumps, and a diffusion constant of 2~\umumps\ 
\cite{HancockHoward99}.  The ratios $\pon/D$ are $0.14$ and $0.12$
respectively, and we use the averaged value $\pon/D=0.13$.  Finally,
based on its molecular weight, we estimated a diffusion constant of $D
= 20$~\umumps\ \cite{NedelecLeibler97}, which yields
$\pon=2.6$~\umumps.

In a tubular, quasi one dimensional geometry, all \mts\ are oriented in
the same direction, (Fig~\ref{FIGIntro}), lower left.  We find that
the steady state profile for the motor concentration is exponential
$f(r) \sim e^{r/a}$ where the distance $a$ is a root of the equation
$a^2 +a/\alpha -1/\beta_1=0$. To relate this
result to the situation of motors in a cell, we now consider a tube
filled with oriented \mts\ connected on one side to a large body.
Motors have a concentration which varies exponentially with the
distance from the cell body.  The concentration at the end of the tube
is $e^{L/a}$ times smaller (or greater depending on the motor sense)
than in the cell body, where $L$ is the length of the tube.  We
estimated $a$ for a cellular extension of a diameter of 2~\um, and for
parameters of the motor kinesin: If the extension contains 20 \mts,
then $a=2$~\um; for the same tube containing a single \mt, $a=30$~\um.
Therefore an unregulated kinesin (which can always bind and move)
would be concentrated even in short extensions of a cell containing
outward polarized \mts.  This is indeed observed \textit{in vivo} for
kinesin heavy chain if it is over-expressed in the absence of the
regulatory light chain \cite{VerheyRapoport98}.

Our experiments were performed in a quasi 2D geometry. We thus
give particular weight to the theoretical analysis of this case which
also presents some interesting theoretical features: Consider firstly
a non-motile binding protein ($v=0$).  At equilibrium, the unbound
molecules are evenly distributed throughout the volume, $\; f(r)={\rm
  const}\; $, while the concentration of bound molecules is
proportional to the local concentration of \mt\ so that $\; b(r) \sim
1/r\; $.  Thus binding of motors induces their accumulation in the
center, where \mts\ are more concentrated.

For general speed $v$ the solutions of the equations in 2D can be
expressed in terms of Whitaker functions, however, simple
analysis (performed by substituting $f(r)\sim r^{-x}$ in eq.~(2))
shows that the solution in the quasi-two dimensional case is well
approximated by power laws beyond the radius $\alpha^3/\beta_2^2$:
$f(r) \sim r^{\alpha/\beta_2}$, and $b(r) \sim \gamma f(r)/r\; .$ Thus
the concentration profile of motors is characterized by an exponent
which is a continuous function of the physical parameters.  From the
above expressions we find $\alpha = \pm 0.8 \mu m $, $1 \mu m <
\beta_2 < 10 \mu m $, and $\gamma = \pm 60 \mu m $. The theory shows that
in large asters (for kinesin, and a sample thickness of $9$~\um, asters
with more than $\sim 600$ \mts), most of the motors are trapped in the center, 
and very few motors are left elsewhere in the sample, effectively causing
a dynamic ``localization''. 
The depletion from the aster center of a kinesin motor moving outwards 
is comparatively weaker. For small asters, motors concentration can be
higher in the center, merely as a consequence of the binding of 
motors to \mts. Total depletion is achieved only for large asters 
(of 1000 \mts), for which outward transport overcomes the pure binding 
effect.

In three dimensions, the perturbation in the concentration due to the
presence of the aster is significant only within a distance 
$\alpha/\sqrt{\beta_3}$ from the center. For large radii, 
$f(r) \sim (1-r_0/r)$. We did not study this situation further due to the absence of experimental results.

\begin{figure}
  \psfig{figure=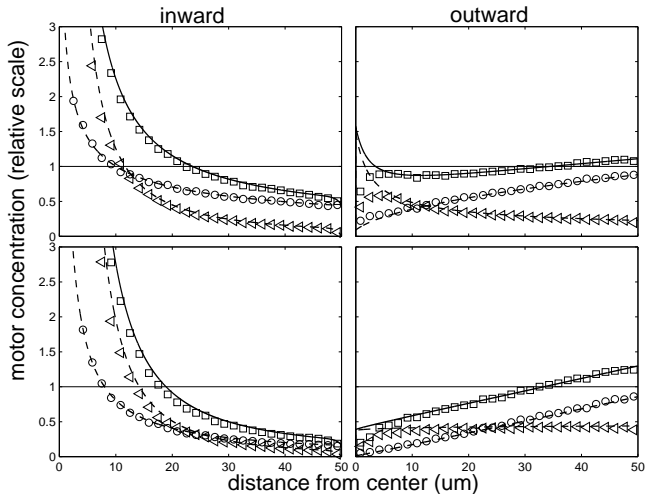,width=8.6cm}
\caption{\label{FIGSim}
  Simulated bound (triangles), free (circles) and total=bound+free
  (squares) motor concentrations and 2D-theory (lines), for kinesin
  moving inward (left) or outward (right) in asters having either 300
  (top) or 600 (bottom) \mts.  As expected some deviations are seen
  near the center of the aster due to finite sample thickness
  implemented in the simulation.  }
\end{figure}

To derive the convection-diffusion equations, we averaged over the
directions transverse to the \mts\ (both angularly, and over the
thickness of the sample). This approximation breaks down
experimentally in a thin sample at the center of a aster where the
geometry is three dimensional, or at large radial distances where the
\mts\ are too far apart for angular averaging.  We thus performed
simulations in a true confined geometry in three dimensions to check
that no substantial errors are introduced in the two dimensional 
theoretical description. In our
simulations the aster is formed by \mts\ of length $L=50$~\um, with
their ``plus'' end in the center of a cylindrical box of radius $L$,
and of thickness 9~\um.  Each motor is characterized by its state
(bound or free), and a vector (position).  To compute the binding of
motors, a rate $p^+$ and an interaction range $\epsilon$ were
introduced, so that at each step, a free motor has a probability
$p^+dt$ to bind to any filament located at a distance (by projection)
closer than $\epsilon$, this effectively corresponds to
$\pon=p^+\pi\epsilon^2$.  At each time step $dt=4.{10}^{-5}$s, bound
motors may detach with a probability $\poff dt$, and otherwise move
radially by a distance $v\, dt$; free motors make random steps, with
$<dx^2>=3D \, dt$.  The motor parameters were taken to mimic kinesin
(see above), with the additional value $\epsilon=50$~\nm\ 
\cite{CoyHoward99}, and $p^+=312$~\ps\ (which yield the correct value
for $\pon$).  Other choices of $\epsilon$ and $p^+$ conserving $\pon$
gave similar curves.  The agreement between simulation and theory,
Fig.  (\ref{FIGSim}) confirms that our analytical approximations are
faithful to the model.

We now turn to our experimental observations on confined quasi two
dimensional samples: Using fluorescent microscopy we measured the
kinesin distribution in asters with the two possible polarities.  These
asters have either the plus or the minus ends of the \mts\ in the center
\cite{NedelecLeibler97,NedelecSurrey00}, and take about 30 minutes to
form.  All data presented here are extracted from two identically
prepared samples, in which many asters of various size formed.  We
measured 115 regular asters with an automatic epi-fluorescence
microscopic setup (Zeiss axioplan 2 with Olympus 100X oil-immersion
objective).  We detected the motors (labeled with the fluorophore
fluorescein) and the \mts\ (labeled with rhodamine) independently.
Digital pictures were taken with a 12-bit CCD camera (Hamamatsu
C4742-95, 1280x1024 pixels).  The camera is linear, and unsaturated
pixel values reflect the relative quantity of protein in the imaged
region.  The sample thickness was $9~\mu m$.

\begin{figure}
  \psfig{figure=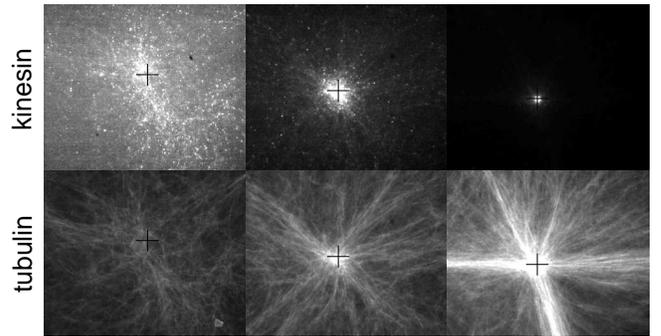,width=8.6cm}
\caption{
  Motor and \mt\ distribution in experimental asters with different
  number of \mts: Fluorescence images (90x70~\um) obtained for the
  motors (top), for the \mts\ (middle).  The cross mark the computed
  measured center.  The intensities of the images are here scaled by
  different factors.  }
\label{FIGExpExamples}
\end{figure}

\begin{figure}
  \center \psfig{figure=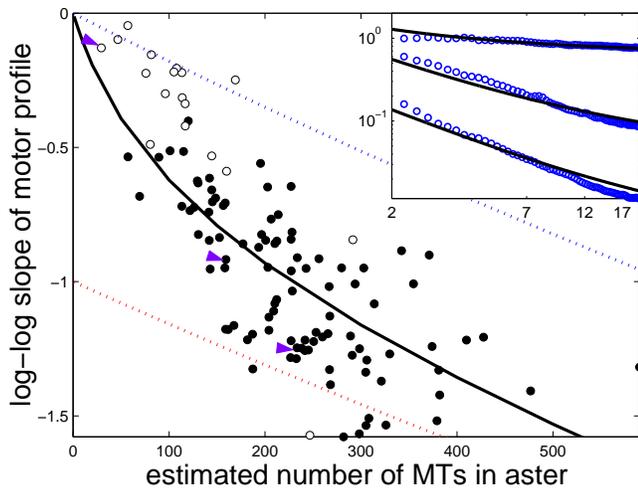,width=8.6cm}
\caption{\label{FIGExExponent}%
  The effective exponent of the kinesin concentration profile becomes
  more negative with the increasing numbers of \mts\ in the aster
  (continuous line: 2D-theory). Filled symbols correspond to regular
  asters for which the radial \mt\ density falls of as $1/r$. Open
  symbols correspond to irregular asters where the density is
  inconsistent with a law in $1/r$.  Arrow heads points to asters
  shown in previous figures.  Inset (log-log): Experimental (circles)
  motor concentration profiles, as extracted from the pictures shown
  in Fig.~\ref{FIGExpExamples}, and 2D-theoretical curves computed for
  the measured number of \mts\ (27, 148, 231) in the aster. }
\end{figure}

The center of the aster and the profiles of fluorescence intensity are
calculated from the image. A common
background pixel value was subtracted from the motor profiles, which
are then normalized. An exponent is obtained by fitting the profile 
in the range 1.5-20~\um\ (the data below 1.5 \um\ is noisy).
To measure the number of \mts\ in the aster, we
fit the profile of \mt\ fluorescence to the function $(M/r+B)$, where
$r$ is the distance from the center.  $B$ is a background, and $M$ is
proportional to the number of \mts.  Calibration was done by manually
counting the \mts\ in five asters.  The $1/r$ profile corresponds to an
homogeneous aster of long \mts.  Experimentally the asters are not
perfect (some, like Fig~\ref{FIGExpExamples}, left are not well
focused).  When the fit of the \mt-profile to $1/r$ is poor, we have no
reason to expect the theory to apply. These asters are plotted with a
different symbol in Fig. (\ref{FIGExExponent}).

We measured the distribution of kinesin in asters containing different
numbers of \mts.  Three typical examples are given in
Fig.~(~\ref{FIGExpExamples}).  Motor profiles of individual asters are
rather well fitted by a power law. They are almost linear on a log-log 
plot, and steeper for bigger asters Fig.~(~\ref{FIGExExponent}), inset.  
Plotting the
exponent of the motor profiles as a function of the number of \mts\
extracted from the \mt\ profiles allows us to compare directly
experiments and theory (see Fig.~\ref{FIGExExponent}).  The data
points, each representing one aster, are scattered around the
theoretical curve, reflecting the heterogeneity of the asters.
However, the major trend in the exponent is correctly predicted by the
theory, so that denser asters are characterized by a larger
localization exponent for the motors.  We also imaged kinesin moving
outwards in asters of normal polarity \cite{NedelecSurrey00}, but the signal was too dim to extract a
reliable profile (the predicted kinesin profile in this situation is
rather flat).

Accumulation or expulsion of molecular motors in asters may have
important functional implications in biology. For instance within a
spindle made of two interacting asters of \mts, minus-ended motors
could concentrate at the poles while plus-ended motors would be
excluded from the same regions.  This could contribute to the
mechanism of spindle assembly, and/or to its mechanical stability.
Interestingly we find that kinesin accumulates in asters of 300-1000
\mts, which is comparable to the number of \mts\ present in spindle
asters of most animal cells. However, the geometry and motors of the
spindle are not the one studied here.

We did not consider the regulation of motor activity: In our study
motors can always bind and move on filaments. Cells use a variety of
processes to counterbalance the impact of motor transport on their
localization.  For example, the folding of kinesin into a non-motile
conformation, in the absence of a cargo
\cite{HackneySuhan92,CoyHoward99bis,StockHackney99} dampens the
transport-induced localization. The recombinant kinesin fraction used
in our experiment lacks this capacity.  Even with this partial
inhibition, the movement of loaded kinesin brings them to places from
which they have to be recycled. Additional regulation mechanisms
include local synthesis and degradation of the motors, involvement of
motors of different directionality transporting each other, etc.
On the other hand, we can also imagine situations in which the unregulated
localization of a motors resulting from their movement can have interesting
consequences.

In summary, motor movements on \mts\ can effectively cause their
``compartmentalization''. The theory provides a full understanding of
the influence of all motor kinetic parameters, and of the geometric
properties of the \mt\ array. 
 
{ \small We thank A.~Ajdari, S.~Blandin, A.~Desai, E.~Karsenti and
  S.~Leibler. }


\begin{thebibliography}{10}
\bibitem{footnote} Experimentally, we always observe formation of
  aggregates (containing kinesin) at the center of the aster,
  reflecting the finite solubility of the protein.
\end{thebibliography}
\end{document}